\documentclass[12pt]{article}
\usepackage{cite}
\usepackage{amssymb}
\usepackage{amsmath}
\usepackage{epsf,graphicx}

\newcommand\ZZ{\hbox{\zfont Z\kern-.4emZ}}
\font\zfont = cmss10 

\newcommand{\F}{{\cal F}_{\chi=1}}
\newcommand{\FF}{{\cal F}_{\chi=2}}

\newcommand{\nc}{N_c}
\newcommand{\nf}{N_f}

\newcommand{\Xt}{\tilde{X}}
\newcommand{\Nt}{\tilde{N}}

\newcommand{\ncbar}{{\Nt_c}}
\newcommand{\nfbar}{{\Nt_f}}
\newcommand{\ut}{\tilde{u}}

\begin{document}

\begin{titlepage}
\begin{flushright}
{\tt hep-th/0312054} \\
UPR-1058-T \\
UW/PT-03-32 \\
INT-PUB 03-21\\
\end{flushright}

\vskip.5cm
\begin{center}
{\huge{\bf Matrix Models, Monopoles and Modified Moduli}}
\vskip.2cm
\end{center}
\vskip0.2cm

\begin{center}
{\sc Joshua Erlich}$^a$, {\sc Sungho Hong}$^{a,b}$, and {\sc Mithat \"Unsal}$^a$
\end{center}
\begin{center}
\vspace*{0.1cm}
$^a${\it Department of Physics, Box 1560, University of Washington, Seattle, WA
98195, USA } \\
$^b${\it Department of Physics and Astronomy, University of Pennsylvania, Philadelphia, PA 19104, USA}

\vspace*{0.1cm}
{\tt erlich@phys.washington.edu, shhong@sas.upenn.edu,
mithat@phys.washington.edu}
\end{center}

\vglue 0.3truecm

\begin{abstract} \vskip 3pt \noindent 
Motivated by the Dijkgraaf-Vafa correspondence,  
we consider the matrix model duals of
${\cal N}$=1 supersymmetric SU$(N_c)$
gauge theories with $N_f$ flavors.  We 
demonstrate via the matrix model
solutions a relation between vacua of theories with
different numbers of colors and flavors.  This relation is due to an 
${\cal N}$=2 nonrenormalization theorem which is inherited by these 
${\cal N}$=1
theories.
Specializing to the case $N_f=N_c$, the simplest theory containing
baryons, we demonstrate that the explicit matrix
model predictions for the locations on the Coulomb branch at which monopoles
condense are consistent with the quantum modified
constraints on the moduli in the theory.  The matrix model solutions include
the case that baryons obtain vacuum expectation values.  In specific cases we
check explicitly that these results are also consistent with the factorization
of corresponding Seiberg-Witten curves.  
Certain results are easily understood
in terms of M5-brane constructions of these gauge theories.

\end{abstract}

\end{titlepage}

\newpage


\section{Introduction} \label{sec:intro} \setcounter{equation}{0}
\setcounter{footnote}{0}
\setcounter{equation}{0}

There has been a recent resurgence of interest in supersymmetric (SUSY) gauge
theories, due in part to the observation of Dijkgraaf and Vafa \cite{DV}
that certain classes of supersymmetric theories can be described by matrix
models.  This correspondence exists for pure ${\cal N}$=2 SUSY U$(N)$ gauge
theory broken to ${\cal N}$=1 by a superpotential $W=\sum_k
g_k\,{\rm Tr}\,\Phi^k$ for
adjoint chiral multiplet $\Phi$, as well as for
other gauge groups \cite{othergauge}, with the addition of matter 
\cite{matter,baryons},
with multi-trace superpotentials \cite{multitrace} and for deformations
of ${\cal N}$=4 gauge theories \cite{tim}.  The
types of results that can be derived or verified from the matrix model
dual include effective glueball superpotentials \cite{DV}, Seiberg duality
\cite{Feng}, and the Affleck-Dine-Seiberg
superpotential at low energies \cite{matter,Demasure}.  Several techniques have 
been developed along these lines.  The low energy superpotential after 
integrating out all of the matter can be derived in some cases, including 
nonperturbative effects, from just planar diagrams in the matrix model 
\cite{DV}.  A new type of duality emerged by relating weakly coupled
theories upon varying the tree level couplings so the theory passes through
strong coupling \cite{newduality}.
Similar results have been derived through consistency with the 
Konishi anomaly \cite{konishi},
and these results are also expected to be consistent with the 
picture of monopole condensation as described by Seiberg and Witten 
\cite{SW1,SW2,DJ,KS}.  It is the last of these points 
of view that we consider in 
this paper.  We will use the matrix model solutions to
relate vacua of different gauge theories.  In some sense this is a type of duality, although
it follows naturally from an ${\cal N}$=2 nonrenormalization theorem which
is inherited by the ${\cal N}$=1 theory obtained by deforming the ${\cal N}$=2 
theory.
We will also study the matrix model
solution for
the $\nf=\nc$ theory in detail and demonstrate explicitly that the location in moduli space
where monopole condensation takes place is consistent with the quantum modified constraint
and factorization of Seiberg-Witten curves in {\em all} vacua of the theory, despite the 
complications due to the presence of baryons.

There is a classic picture of confinement as arising from the dual 
Meissner effect associated with the condensation of monopoles 
\cite{confinement}.  
Just as superconductors confine 
magnetic fields when electrons condense, gauge theories are 
expected to confine 
electric fields when magnetic 
monopoles condense.  While there is some evidence for 
this picture in ordinary QCD on the lattice \cite{lattice}, the 
occurrence of monopole condensation with confinement is better understood in 
certain supersymmetric theories \cite{SW1}.  

In particular, consider 
${\cal N}$=2 supersymmetric SU$(N_c)$ Yang-Mills theory with $N_f$ fundamental
massless flavors.
The vacuum structure of these theories was
studied in detail in \cite{APS}.  The moduli
space of vacua is split into Higgs branches, in which the matter scalars
$q_i$ have vacuum expectation values (vevs), and the Coulomb branch, in which
the adjoint scalars $\phi^a$, $a=1,\dots,N_c^2-1$ have vevs.  
At generic points in the moduli space $q_i$ and $\phi^a$ do not both have vevs together.  
The exception is along special submanifolds
in the moduli space, called roots, 
where the Higgs and Coulomb branches meet.  
The roots are themselves classified by the gauge symmetry in the vacuum.
There are non-baryonic roots in which the unbroken gauge symmetry is
SU$(r)\times$U$(1)^{N-r}$ with $r\leq[N_f/2]$, and if $N_f\geq N_c$ there is also
a baryonic root in which the gauge symmetry is SU$(N_f-N_c)\times$U$(1)^{
2N_c-N_f}$.  (If $N_f$=$N_c$ then the unbroken
gauge symmetry at the baryonic root
is just U$(1)^{N_c-1}$.)  At generic points on the Coulomb branch
the unbroken gauge symmetry is U$(1)^{N_c-1}$, so the roots typically
have an enhanced gauge symmetry if the quarks are massless.

Furthermore, the roots contain special points or submanifolds on which 
monopoles or dyons 
(which we will refer to collectively as monopoles) become 
massless.  
As we will review,
upon breaking to ${\cal N}$=1 supersymmetry the only vacua that
survive are those points on the roots of the moduli space
in which each of the unbroken
U$(1)$'s has monopoles charged under it which then condense giving rise to
confinement of electric charges as discussed above. 
These points typically occur at strong coupling, and the 
weak coupling degrees of freedom, $\Phi\equiv \sum_{a=1}^{N_c^2-1}\Phi^a\,T^a$
with $T^a$ generators of SU$(N_c)$ in the fundamental representation, 
are not a good description 
of the theory near these points.  Instead, there are weakly coupled
dual degrees of freedom, $\Phi_D\equiv \sum_{a=1}^{N_c^2-1}\Phi_D^a\,T^a$, 
which are defined at each point in the moduli space 
up to gauge equivalences.

The $D$-term equations of motion $[\Phi,\Phi^\dagger]=0$ 
force the adjoint to belong to the Cartan 
subalgebra of SU$(N_c)$ so that the
adjoint scalars $\phi_D$ can be diagonalized, and the $N_c-1$ independent
diagonal elements form a basis of adjoints of the unbroken U(1)$^{N_c-1}$ 
gauge group
along the Coulomb branch.  We will concentrate on the theory with $N_f$=$N_c$,
and we will be most interested in those roots of the Higgs branches
at which monopoles $Q_{mon}^i$ condense.  At the baryonic root 
$N_c-1$ of these monopoles
are each charged under one of the unbroken U(1)'s, and one of the monopoles,
$Q_{mon}^N$, has the opposite charge under each of the U(1)'s.
The monopoles form ${\cal N}$=2 
hypermultiplets, so for each ${\cal N}$=1 chiral multiplet $Q_{mon}^i$ there
is a charge conjugated chiral multiplet $\tilde{Q}_{mon}^i$.
It is convenient to collect the monopoles into a fundamental of the high
energy SU$(N)$ gauge group, $Q_{mon}$ and $\tilde{Q}_{mon}$. 
The monopoles acquire the ${\cal N}$=2 
superpotential, \begin{equation}
W_{mon}={\rm Tr}\,Q_{mon}\,\Phi_D\,\widetilde{Q}_{mon},
\end{equation}  
where the trace is over the SU$(N_c)$ gauge group.\footnote{Note that the normalization
of the superpotential $W_{mon}$ differs from the usual ${\cal N}$=2 normalization by a factor
of $1/\sqrt{2}$, but was chosen for consistency with reference 
\cite{Demasure,DJ} which we refer
to often in this paper.}  The monopoles are
massless when the vev of $\Phi_D$ vanishes.

The adjoint scalars $\phi$ (or $\phi_D$) are not gauge invariant: $\phi$ lies
in the Cartan subalgebra of SU$(N_c)$, and there are elements of SU$(N_c)$/U$
(1)^{N_c-1}$ which leave $\phi$ in the Cartan subalgebra but permute the
diagonalized elements of $\phi$.  Instead, it is convenient to parametrize the
Coulomb branch by gauge invariant polynomials \cite{LT},
\begin{equation}
u_k=\frac{1}{k}\,{\rm Tr}\,\Phi^k, \ \ \ k=2,\dots,N. \end{equation}

If ${\cal N}$=2 supersymmetric mass terms of the
form $\sum_{i=1}^{N_f}m_i {\rm Tr}\,Q_i\tilde{Q}_i$
are added to the superpotential then generically the
Higgs branch will be completely lifted.  From the point of view of
the classical theory the reason is that the combination of
$D$-term and adjoint $F$-term equations of motion force 
the squark vevs, written as $N_c\times N_f$ matrices (with $N_f=N_c$ here) 
to take either the non-baryonic form \cite{APS}:
\begin{equation}
Q=\left(\begin{array}{cccccc}
q_1&      &           &   &   &   \\
   &\ddots&           &   &   &   \\
   &      &q_{[\nf/2]}&   &   &   \\ 
   &      &           & 0 &   &   \\
   &      &           &   &\ddots  &   \\
   &      &           &   &   & 0 \end{array}\right), \hspace{1cm}
\tilde{Q}^T=\left(\begin{array}{cccccc}
 0  &  &    &q_1&      &   \\
   & \ddots  &    &   &\ddots&   \\
   &   &   0 &   &      &q_{[\nf/2]}\\ 
   &   &    &   &      &   \\
   &   &    &   &      &   \\
   &   &    &   &      &   \end{array}\right) \end{equation}
or the baryonic form, \begin{equation}
Q=\left(\begin{array}{cccccc}
q  &      &           &   &   &   \\
   &q   &           &   &   &   \\
   &      & \ddots &   &   &   \\ 
   &      &           & \ddots &   &   \\
   &      &           &   &\ddots  &   \\
   &      &           &   &   &  q \end{array}\right), \hspace{1cm}
\tilde{Q}^T=\left(\begin{array}{cccccc}
 \tilde{q}  &  &    &   &      &   \\
   & \tilde{q}  &    &   & &   \\
   &   & \ddots  &   &      & \\ 
   &   &    &\ddots  &     &   \\
   &   &    &   & \ddots   &   \\
   &   &    &   &      &  \tilde{q} \end{array}\right). \end{equation}

In components the matter $F$-term equations are: \begin{eqnarray}
Q_a^i\Phi_b^a &=& m_i Q_b^i \\
\Phi_b^a\tilde{Q}^b_i &=& m_i \tilde{Q}^a_i. \end{eqnarray}
With $\Phi$ diagonalized, it is clear that there are only nonbaryonic
solutions with nonvanishing $Q$ if pairs of masses are equal, and there are only
baryonic solutions if $\sum_i m_i=0$ (for tracelessness of $\Phi$).  
While we only considered the case $N_f=N_c$ above, 
the general result remains true for $N_f<N_c$: there
are only non-baryonic Higgs branches if pairs of quark masses are equal.
As we will see in Section~\ref{sec:M} this behavior is
immediately apparent from the Type IIA and M-theory brane constructions 
of these theories.  

If the ${\cal N}$=2 theory is weakly broken to ${\cal N}$=1 SUSY
with a superpotential of the form \begin{equation}
W_{{\cal N}=1}=\sum_k g_k\,u_k, \label{eq:W1}\end{equation} 
then the effective superpotential including the light monopoles is 
$W_{{\cal N}=1}+W_{mon}$.  At the baryonic branch in the $N_f=N_c$ theory there
are no additional massless fields besides the monopoles.

Integrating out the $u_k$ leads to, \begin{equation}
{\rm Tr}\,Q_{mon}\,\frac{\partial\Phi_D}{\partial 
u_k}\,\tilde{Q}_{mon}=-g_k, \label{eq:mon1}
\end{equation}
which implies that, as long as $\partial\Phi_D/\partial u_k$ is generic
and nonvanishing, at least $N_c-1$ of the $N_c$
monopoles acquire vacuum expectation values.  Then 
the equations of motion for the monopoles set, \begin{equation} 
\Phi_D(u_k)=0, \label{eq:mon2}\end{equation} 
so the ${\cal N}$=1 vacua are stuck to the points at which the monopoles 
are massless, and they condense via (\ref{eq:mon1}).  In the massless ${\cal N}$=2
theory these points
are at the roots of the Higgs branches, but for generic masses such points continue to
exist despite the absence of Higgs branches.
However, the Coulomb branch moduli at the ${\cal N}$=1
vacua vary smoothly as the quark masses vary, so we will continue to label a vacuum of the
${\cal N}$=1 theory as baryonic or nonbaryonic depending on which Higgs branch the
vacuum flows to as the quark masses are taken to zero.

The low energy superpotential is then
\begin{equation}
W_{eff}=W_{{\cal N}=1}(u_k^0), \label{eq:Weffmon}\end{equation}
where $u_k^{(0)}$ are the solutions of (\ref{eq:mon2}).

If the quarks are massless, then at
the non-baryonic roots the gauge symmetry is enhanced to SU$(r)
\times$U(1)$^{N_c-r}$ 
with $r\leq N_f/2$.  In addition to the massless monopoles
there are $N_f$ hypermultiplets in the fundamental of SU$(r)$.  If
pairs of quark masses are equal but otherwise generic, then non-baryonic
branches 
remain unlifted, but the SU$(r)$ factor is broken to U(1)$^{r-1}$ and there
is one hypermultiplet charged under each of these U(1) factors.
In any case, these light fields
also acquire the ${\cal N}$=2 superpotential $\sum_i Q_i\Phi\tilde{Q}_i$, and
the $Q_i$ equations of motion then set this term in the superpotential to
zero as for the monopoles.  The effective superpotential in the vacuum is then
given by (\ref{eq:Weffmon}), just as in the baryonic vacuum.    


If the adjoint mass ($g_2$ in the tree level superpotential) is large
compared to the dynamical scale $\Lambda$, then
the adjoint can be integrated out at high energies where the electric
theory is weakly coupled.  The description of the theory in terms
of quarks is valid at that scale, and running down from there
the mesons and baryons provide a valid
low energy description of the theory. This description of the low energy 
theory should yield the
same low energy superpotential as the monopole description (\ref{eq:Weffmon}),
assuming there is no phase transition as the adjoint mass is varied.
Including flavor masses, the tree level superpotential is,
\begin{equation}
W_{tree}=\sum_{i=1}^{N_f} {\rm Tr}\,(Q_i\Phi\tilde{Q}_i+m_iQ_i\tilde{Q}_i)+
\sum_{k=2}^{N_c} g_k u_k, \label{eq:Wtree}\end{equation} 
where the first sum is the usual ${\cal N}$=2 superpotential and the second is
the ${\cal N}$=2 breaking superpotential.  In the $N_f$=$N_c$ theory there
is a quantum modified constraint among the mesons and baryons.
We will check that the matrix model prediction of the low energy superpotential
with only $g_2$ nonvanishing
agrees with the field theory prediction from (\ref{eq:Wtree}) and the
quantum modified constraint, and we
will check in specific cases that these descriptions correctly
predict the locations on the Coulomb branch 
at which monopole condensation occurs.

In Section \ref{sec:Nf=Nc} we review the explicit
matrix model results of Demasure and Janik \cite{DJ}
and we demonstrate that the extension of these results 
to the case $N_f=N_c$ correctly reproduces the nonperturbative field theory
predictions for the low energy superpotential at the appropriate vacua, with
the expected quantum modified constraints on the moduli. 
In Section
\ref{sec:SU2} we discuss the SU(2) and SU(3) theories in more detail, 
and demonstrate explicitly
that the corresponding Seiberg-Witten curve factorizes as required for 
consistency with the picture of monopole condensation. 
In Section~\ref{sec:BvsNB} 
we use the matrix model solution to
demonstrate a relation between vacua of theories with different numbers
of colors and flavors.
We elucidate some of these results via  M5-brane 
constructions in Section~\ref{sec:M}.
We conclude in Section 
\ref{sec:Conclusions}.

\section{Matrix model predictions for $N_f\leq N_c$}
\label{sec:Nf=Nc}
\setcounter{equation}{0}
In this section we summarize the matrix model results of Demasure \& Janik 
\cite{DJ} for SU$(N_c)$ with $N_f<N_c$ and we discuss the 
extension of these results to the case $N_f=N_c$.

\subsection{$N_f<N_c$}
Dijkgraaf and Vafa \cite{DV} suggested that ${\cal N}$=2 supersymmetric gauge
theories perturbed by single-trace superpotentials of the form $W_{{\cal N}=1}=
\sum_{k=2}^{N_c} 
g_k \,{\rm Tr}\,\Phi^k$ have an equivalent description in terms of
a large-$N$ 
matrix model whose action is simply \begin{equation}
V(\Phi,Q_i,\tilde{Q}_i)=N/S \left({\rm Tr}\, W_{{\cal N}=1}(\Phi)+ 
\sum_i \left(m_i
Q_i \tilde{Q}_i+Q_{i}\Phi\tilde{Q}_i\right)\right), 
\label{eq:matrixpot}
\end{equation}
for the $N\times N$ matrix
$\Phi$, $1\times N$ vector $Q_i$ and $N\times 1$ vector
$\tilde{Q}_i$. Another way to think of confinement 
below the scale of the flavor masses is via gaugino condensation in the
${\cal N}$=1 pure Yang-Mills theory.  Again, the various descriptions of
the theory (in terms of quarks, monopoles or glueballs), are
valid in different regimes of mass parameters or energies, and the assumption
is that there is no phase transition upon varying these parameters so that
the resulting low energy descriptions should be equivalent.
In (\ref{eq:matrixpot}) $S$ will be interpreted
as the glueball superfield in the gauge theory.  
In particular, the leading $N$ contributions
to the matrix model partition function are interpreted in terms of the 
effective glueball superpotential in the gauge theory.  If the glueball
superpotential is written, \begin{equation}
W_{eff}(S)=N_c\frac{\partial{\cal F}_{\chi=2}(S)}{\partial S}+
{\cal F}_{\chi=1}, \label{eq:WS}\end{equation}
then the matrix model prediction for ${\cal F}_{\chi=1}$ and 
${\cal F}_{\chi=2}$ is, 
\begin{equation}
e^{-\frac{N^2}{S^2}\FF(S)-\frac{N}{S}\F(S)+{\cal O}(N^0)}=
\int D\Phi\,DQ_i\,D\tilde{Q}_i\,e^{-\frac{N}{S}\left({\rm Tr}\, 
W_{{\cal N}=1}(\Phi)+ 
\sum_im_i Q_i \tilde{Q}_i+Q_{i}\Phi\tilde{Q}_i\right)}. \label{eq:MM}
\end{equation}
The glueball is expected by symmetry arguments 
\cite{VY} to 
couple to $\log\Lambda^{2N_c-N_f}$ where $\Lambda$ is the dynamical scale of
the theory.  Integrating in the glueball $S$ via a Legendre transform
of the low energy superpotential (\ref{eq:Weffmon}) with respect to
$\log\Lambda^{2N_c-N_f}$ \cite{integrating-in}, the effective glueball superpotential
derived from the matrix model should be equated with \cite{DJ},
\begin{equation}
W_{eff}(S;\Omega,\Lambda)=S\log(\Lambda/\Omega)^{2N_c-N_f}+\sum_k g_k
u_k^0 (\Omega). \label{eq:S2}\end{equation}
In (\ref{eq:S2}) 
the moduli are to be understood as functions of the quark masses 
and the scale $\Omega$, which is set
to the dynamical scale $\Lambda$ by the $S$ equation of motion.
By integrating out the $\Omega$ auxiliary field, one obtains the low energy 
effective potential as a function of $S$.  
One can easily check that integrating out
the glueball $S$ reproduces the superpotential $W_{{\cal N}=1}(u_k^0)$.

By explicitly
calculating the matrix model integrals in (\ref{eq:MM}), determining
the glueball superpotential (\ref{eq:WS}) and comparing with (\ref{eq:S2}), 
Demasure and Janik
calculated the values of the moduli $u_k^0$ in the vacua of the
theory as a function of the
dynamical scale $\Lambda$ and the flavor masses $m_i$.
In fact, they were only interested in the baryonic vacua, so
they selected a subset of the matrix model solutions.  There is a
$\mathbb{Z}_2^{N_f}$
ambiguity in the branch choice of a 
square root that appears in the identification of the effective superpotential
in terms of matrix model variables.  This can be seen either via the 
breakdown of the perturbative expansion of the
matrix model free energies ${\cal F}_{\chi}$ at the branch point, or
via the existence of
multiple saddle point solutions in the matrix model expressed in terms of
mesons as opposed to quarks \cite{Ohta}.
By choosing
the other branches we recover all of the vacua in the gauge theory.  In the case
that Higgs branches are unlifted there
is a simple relationship between the various branches of the
square roots and the Higgs branches.

Without repeating the details of the matrix model integrals, we quote the
results in the form given in \cite{DJ}.
The result of the matrix model calculation for the moduli $u_k$ 
in the ${\cal N}$=1
vacua is given implicitly in terms of the dynamical scale $\Lambda$
of the high energy theory, and $u_1={\rm Tr}\,\Phi$, which is set to zero
for the SU$(N_c)$ theory.  Following \cite{DJ}, we define the matrix model
parameters $T$ and $R$ in terms of the field theory parameters $u_1={\rm Tr}\,
\Phi$ and $\Lambda$ via,
\begin{multline}\label{EQu1}
  u_1 ( R, T, m_i ) = N_c T - \sum_{i = 1}^{n_-} \frac{1}{2} \left( m_i + T -
  \sqrt{( m_i + T )^2 - 4 R} \right)\\
   - \sum_{i = n_- + 1}^{\nf} \frac{1}{2}
  \left( m_i + T + \sqrt{( m_i + T )^2 - 4 R} \right),
\end{multline}
\begin{multline}\label{EQLambda}
  \Lambda^{2N_c-N_f} = R^{\nc - \nf} \prod_{i = 1}^{n_-} 
\frac{1}{2} \left( m_i +
  T - \sqrt{( m_i + T )^2 - 4 R} \right) \\
   \times \prod_{i = n_- + 1}^{\nf} \frac{1}{2}
  \left( m_i + T + \sqrt{( m_i + T )^2 - 4 R} \right).
\end{multline}
Then, generalizing the result of Demasure and Janik \cite{DJ} to include the
various branches of the matrix integrals (see also \cite{Ohta}), the
moduli in the ${\cal N}$=1 vacua are given by: 
\begin{multline}
u_p^{0}=N_c\,{\cal U}_p^{pure}(R,T)+\sum_{i=1}^{n_-}{\cal U}_p^{-\,matter}(
R,T,m_i) 
+\sum_{i=n_-+1}^{N_f}{\cal U}_p^{+\,matter}(R,T,m_i)
, \label{eq:upfact} \end{multline}
where \begin{eqnarray}
{\cal U}_p^{pure}(R,T)&=&\frac{1}{p}\sum_{q=0}^{[p/2]}\left(\begin{array}{c}
p \\ 2q \end{array}\right)\left(\begin{array}{c}2q \\ p\end{array}\right)
R^qT^{p-2q} \nonumber \\
{\cal U}_{p\geq 2}^{+\,matter}(R,T,m)&=&\sum_{n=0}^{p-2} c_{p,n} Rf_n(z^+)-
\frac{v_p}{2}\left(m+T+\sqrt{(m+T)^2-4R}\right) \label{eq:Up}\\
{\cal U}_{p\geq 2}^{-\,matter}(R,T,m)&=&\sum_{n=0}^{p-2} c_{p,n} Rf_n(z^-)-
\frac{v_p}{2}\left(m+T-\sqrt{(m+T)^2-4R}\right). \nonumber\end{eqnarray}
$c_{p,n}$ and $v_p$ are functions of $R$ and $T$, and are given by \cite{DJ},
\begin{eqnarray}
c_{p,n}&=&2^nR^{n/2}\sum_{k=0}^{\left[\frac{p-n-2}{2}\right]}
\left(\begin{array}{c}2k \\ k \end{array}\right)
\left(\begin{array}{c}p-1 \\ 2k+n+1 \end{array}\right)
R^k\,T^{p-n-2-2k}\\
v_p&=&\sum_{q=0}^{[p/2]}\frac{p-2q}{p}
\left(\begin{array}{c}p \\ 2q \end{array}\right)
\left(\begin{array}{c}2q \\ q \end{array}\right)
R^q\,T^{p-2q-1}.
\end{eqnarray}
The $f_n(z^{\pm})$ are polynomials determined in
\cite{DJ}, up to the aforementioned branch choice which appears again in
$z^{\pm}(R,T,m)$: \begin{eqnarray}
z^\pm&=&\frac{m+T}{2R}\left(m+T\mp\sqrt{(m+T)^2-4R}\right)
\label{eq:z}
\end{eqnarray}
The odd-looking choice of positive and negative roots in (\ref{eq:z}) was 
made for consistency with (\ref{eq:Up}).  
The $r$th Higgs branch 
corresponds to all solutions with $N_f-|N_f-2n_-|=r$, where $n_-$ is 
the number of $z^-$'s in the solution (\ref{EQu1})-(\ref{eq:upfact}).
The reflection of all $z^+$'s and
$z^-$'s, i.e. taking $n_-\rightarrow N_f-n_-$, corresponds to solutions in 
the
same $r$th Higgs branch so we will generically take $n_-\leq [N_f/2]$ and
call $n_-=r$ with
the understanding that we must also consider solutions with $n^-
\rightarrow N_f-n_-$ in the same Higgs branch.

The choice $r=0$ corresponds to the
baryonic branch, and the other roots correspond to the non-baryonic branches.
Recall also that we continue to label the vacua as baryonic or nonbaryonic despite the
fact that for generic masses they are unrelated to Higgs branches of the ${\cal N}$=2 theory.

For simplicity we will concentrate on tree level
superpotentials containing only $u_2$ and $u_3$,
so from (\ref{eq:Up})
we will only need $f_0(z)$ and $f_1(z)$, which are \cite{DJ},
\begin{eqnarray}
f_0(z)&=&\frac{1}{2(z-1)}\\
f_1(z)&=&\frac{3z-4}{6(z-1)^{3/2}}. \end{eqnarray}
This completes the summary of matrix model results we will need, and for
the complete discussion we refer the reader to \cite{DJ}.

\subsection{$N_f=N_c$}
Here we study the SU$(N_c)$ theory with $N_c$ flavors.  While the matrix
model approach has additional complications when baryons are present 
\cite{baryons}, we will see that the baryons vanish in all vacua except when the sum of
the quark masses vanishes.  Because the values of the
moduli at the points where monopoles become
massless vary smoothly with the quark masses, the effective superpotential 
is also a smooth function of the quark masses.  Hence, the matrix
model predictions in previous sections for the low energy superpotential, which
assumed that the baryons vanish,
must correctly predict the effective superpotential on 
the baryonic Higgs branches as well.
This allows us to easily extend the results of \cite{Demasure,DJ} to the case $N_f=N_c$.
In fact, it has been argued that the quantum modified constraint 
including the baryons $(\det X-B\tilde{B}=\Lambda_{low}^{2N})$
follows from the matrix model with sources for the
baryons included in the matrix model \cite{baryons}. 
We will check here that by assuming that the baryons vanish,
the matrix model predictions for the moduli $u_k^0$ 
are consistent with a field theory 
analysis, and in Section~\ref{sec:SU2} we will demonstrate that these
results correctly
predict factorization conditions for related Seiberg-Witten curves.

As in the theories with $N_f<N_c$ studied in \cite{DJ},
$z^\pm(R,T,m_i)$ will be related to the nonvanishing components of the
diagonalized $N_f\times N_f$ meson matrix, $X_{ij}=M\tilde{X}_i\delta_{ij}$.  
More precisely,
\begin{equation}
\tilde{X}_{i}\equiv\frac{R}{m+T}\,z^\pm(R,T,m_i),
\label{eq:Xii}\end{equation}
where the sign choice is the same as that which determines the choice of 
vacuum in (\ref{EQu1}) and (\ref{EQLambda}), for example.
We immediately recognize
(\ref{EQLambda}) as the quantum modified constraint of the ${\cal N}$=1
theory with $N_f=N_c$ when the baryons vanish, \begin{equation}
\det\,X_{ij}=\Lambda^{N_c}M^{N_c}, \label{eq:QMC}\end{equation}  
where the 
one-loop matching condition determines the low energy dynamical
scale below the mass $M$ as $\Lambda_{low}^{2N_c} =\Lambda^{N_c}M^{N_c}$.
We would like to demonstrate that the low energy effective superpotential
at the various vacua agree with the matrix model prediction.  
Recall that 
the gauge theory at low energies has dual descriptions, either in terms
of mesons and baryons 
or in terms of monopoles.  
We discussed the description
in terms of monopoles in the Introduction.  If the only ${\cal N}$=2
breaking term in the action is an adjoint mass, then
the effective
superpotential in the vacuum of the theory with 
$W_{{\cal N}=1}=M/2\,{\rm Tr}\,\Phi^2=Mu_2$ is,
\begin{equation}
\label{eq:WM}
W_{eff}=M\,u_2^{0},\end{equation}
where $u_2^0$ is evaluated at a solution of (\ref{eq:mon2}).
We obtain the description
in terms of mesons and baryons by taking the ${\cal N}$=2 superpotential 
deformed by the superpotential (\ref{eq:W1}), integrating out the
adjoint and the massive matter, and adding the expected
quantum modified constraint by hand.  This is expected to be valid when the
adjoint mass $M$ is much larger than the dynamical scale $\Lambda$
(so that the theory is weakly coupled at $M$), and then also for
smaller values assuming there is no phase transition as the adjoint mass is
varied.  The full tree level superpotential is, 
\begin{equation}
W=\frac{M}{2}{\rm Tr}\,\Phi^2+\sum_{i=1}^{N_f} {\rm Tr}\, Q_i\Phi\tilde{Q}_i
+\sum_{i=1}^{N_f}m_i\,Q_i\tilde{Q}_i+\lambda_\Phi\,{\rm Tr}\,\Phi,
\end{equation}
where $\lambda_\Phi$ is a Lagrange multiplier enforcing the tracelessness
of $\Phi$ in the SU$(N_c)$ theory.  The equations of motion for $\Phi$ are,
\begin{equation}
M\Phi_a^b+\sum_iQ_{ia}\tilde{Q}_i^b+\lambda_\Phi\,\delta_a^b=0, 
\label{eq:PhiEOM}\end{equation}
the trace of which determines $\lambda_\Phi$: \begin{equation}
\lambda_\Phi=-\frac{1}{N}\,\sum_i {\rm Tr}\, Q_i\tilde{Q}_i. \end{equation}
Then we integrate out $\Phi$ using (\ref{eq:PhiEOM}) to obtain, 
\begin{equation}
W_{eff}=-\frac{1}{2M}{\rm Tr}\,X^2+\frac{1}{2N_cM}({\rm Tr}\,X)^2+
{\rm Tr}\,mX, \label{eq:WeffFT}\end{equation}
where the traces are now over flavor indices, and we have defined, 
\begin{eqnarray} X_{ij}&\equiv& \sum_{a=1}^{N_c} Q_{ia}\tilde{Q}_j^a, \\
m_{ij}&\equiv& m_i \delta_{ij}. \end{eqnarray}
The quantum modified constraint can be added to
(\ref{eq:WeffFT}) via a Lagrange multiplier,
$\lambda_X
({\rm det}\,X-\Lambda^{N_c}M^{N_c})/M^{2N_c-1}$.  
The vacua are determined by the equations of motion that follow
from (\ref{eq:WeffFT}) 
together with the quantum modified constraint, 
which are solved by a diagonal meson matrix, 
$X_{ij}=M\tilde{X}_i\delta_{ij}$.
In order to compare the results with the Seiberg-Witten description of the
theory it is convenient to define $\tilde{X}_{ij}\equiv X_{ij}/M$.  Then 
(\ref{eq:WeffFT}) becomes, \begin{equation}
W_{eff}=M\left(-\frac{1}{2}{\rm Tr}\,\tilde{X}^2+\frac{1}{2N_c}({\rm Tr}\,
\tilde{X})^2+{\rm Tr}\,m\tilde{X}+
\lambda_X({\rm det}\,\tilde{X}-\Lambda^{N_c})\right), 
\label{eq:Wefftilde}\end{equation}
and the equations of motion are, \begin{equation}
-\tilde{X}_i+\frac{1}{N}({\rm Tr}\,\tilde{X})+m_i+\frac{\lambda_X}{\tilde{X}_i}
\prod_{k=1}^{N_f}\tilde{X}_k. \label{eq:EOMtilde}\end{equation}

Note that we have implicitly assumed that the baryons vanish in the analysis above.
Including the possibility for baryonic vevs, the Lagrange multiplier term 
in (\ref{eq:Wefftilde})
is modified to $\lambda_X(\det\,\Xt-B\bar{B}-\Lambda^{Nc})$, where with the inclusion
of the Lagrange multiplier we are to treat $B$ and $\bar{B}$ as independent of the mesons
$\Xt$.  Then the $B$ and $\bar{B}$ equations of motion are: \begin{equation}
\lambda_X B=\lambda_X\bar{B}=0. \label{eq:B}\end{equation}
Hence, the baryons can only be nonvanishing if the Lagrange multiplier $\lambda_X$
vanishes.  So let us assume that $\lambda_X=0$.  Then, summing the equations of motion
(\ref{eq:EOMtilde}) over the $N_c$ flavors 
we easily see that a solution exists only if
$\sum_i m_i=0$.  As explained earlier, the fact that such solutions occur only for isolated
regions in the parameter space of the theory, in addition to the fact that the low energy
superpotential varies smoothly with these parameters, implies that the low energy
superpotential is independent of the baryon vevs in the baryonic vacua.  We
will see an example of this in Section~\ref{sec:su2}.

We will now demonstrate that the alternative description for the low energy
superpotential as $W_{eff}=Mu_2^0$, with the 
matrix model prediction for $u_2$ in the vacuum, agrees with 
(\ref{eq:Wefftilde}) and (\ref{eq:EOMtilde}).
From (\ref{eq:upfact}) and (\ref{eq:Up}), we have,
\begin{multline}\label{EQu2}
  u_2^0 ( R, T, m_i ) = \nc (\frac{T^2}{2} + R) \\
  + \frac{1}{4} \sum_{i =
  1}^r \left[( m_i - T ) \left( m_i + T - \sqrt{( m_i + T )^2 - 4 R} \right)
-2R\right] \\
  + \frac{1}{4} \sum_{i = r + 1}^{\nf} \left[( m_i - T ) \left( m_i + T + \sqrt{(
  m_i + T )^2 - 4 R} \right)-2R\right] .
\end{multline}
We can determine $T$ using (\ref{EQu1}) with $u_1$=0 and the 
identification (\ref{eq:Xii}),
from which it follows that, \begin{equation}
T=\frac{1}{N_c}\,{\rm Tr}\,\tilde{X}. \label{eq:TTr}\end{equation}
Then, solving for $R$ from (\ref{eq:Xii}) and tracing over the flavor indices
we obtain,
\begin{equation}
R=\frac{1}{N_c}\left(-{\rm Tr}\,\tilde{X}^2+\frac{1}{N_c}({\rm Tr}\,
\tilde{X})^2+{\rm Tr}\,m\tilde{X}\right). \end{equation}
Substituting these relations for the matrix model variables
$R$ and $T$ into (\ref{EQu2}) gives the matrix model prediction for
the effective superpotential $W_{eff}=Mu_2^0$ in terms of the mesons $\Xt_{ij}$
at the vacua: \begin{equation}
W_{eff}^{vac}=M\left(-\frac{1}{2}\,{\rm Tr}\,\tilde{X}^2+\frac{1}{2N_c}
({\rm Tr}\,\tilde{X})^2+{\rm Tr}\,m\tilde{X}\right), \label{eq:Weffmatrix}
\end{equation}
which,
together with the quantum modified constraint (\ref{eq:QMC}), 
exactly matches the field theory result (\ref{eq:Wefftilde}).  It
remains to derive the equations of motion (\ref{eq:EOMtilde}) from the matrix 
model.

Solving for $R$ 
from the definition of $\tilde{X}_i$ in the matrix model (\ref{eq:Xii})
we have, 
\begin{equation}
R=-\Xt_i^2+(m_i+T)\Xt_i. \label{eq:R}\end{equation}
Recalling (\ref{eq:TTr}), if $\Xt_i\neq 0$
(\ref{eq:R}) can be rewritten, \begin{equation}
-\Xt_i+\frac{1}{N_c}\sum_i\Xt_i+m_i-\frac{R}{\Xt_i}. \end{equation}
Comparing with the equations of motion (\ref{eq:EOMtilde}), we see that the
matrix model predicts the expected equations of motion and identifies the
Lagrange multiplier with, \begin{equation}
\lambda_X=-\frac{R}{\prod_k\Xt_k}=-\frac{R}{\Lambda^{N_c}}, \end{equation}
where in the last step we used the quantum modified constraint (assuming
the baryons vanish), which we have
already seen follows from the matrix model.  

This completes the demonstration
that the matrix model prediction for the low energy superpotential in the
$N_f=N_c$ theory agrees with the field theory.  However, we would like to 
stress that the matrix model does more.  If the two equations 
(\ref{EQu1}) and (\ref{EQLambda}) defining the
matrix model parameters $R$ and $T$ can be solved, then the values of all of
the moduli $u_p^0$ are determined by the matrix model in terms of those
two parameters.  This follows from an analysis similar to that above
with a generic superpotential $\sum_k g_k\,{\rm Tr}\,\Phi^k$ in place of
the mass term $M/2\,{\rm Tr}\,\Phi^2$.
In the following we explicitly check in certain cases
that the above solutions factorize the appropriate
Seiberg-Witten curves.

\section{Factorization of Seiberg-Witten curves}
\label{sec:SU2}
\setcounter{equation}{0}

The gauge theories discussed above would have ${\cal N}$=2
supersymmetry were it not for the tree level superpotential
$W_{{\cal N}=1}$.  The low energy dynamics of these theories on the Coulomb 
branch has a solution
in terms of a Seiberg-Witten curve \cite{SW1,SW2}.  The ${\cal N}$=2
prepotential of the low energy effective theory on the Coulomb branch
is determined in terms of integrals of certain one-forms over the
cycles of the Seiberg-Witten curve.  In addition, the spectrum of
monopoles and dyons is also determined by such integrals.  As
discussed earlier, when the theory is weakly broken to ${\cal N}$=1
supersymmetry by the tree level superpotential the Coulomb branch is lifted
except for those points at which monopoles would be massless, and these 
monopoles condense.  There is not expected to be a phase transition as the
tree level couplings are varied, so the monopoles are expected to condense
also when the theory is strongly broken to ${\cal N}$=1.  This idea is tested
by the matrix model description of the theory.
In particular, the matrix model predictions for the moduli at the
ground states of the theory are supposed to correspond to 
the points at which the 
monopoles of the ${\cal N}$=2 theory would be massless.

\begin{figure}[t]
\epsfxsize=3in
\centerline{\epsfbox{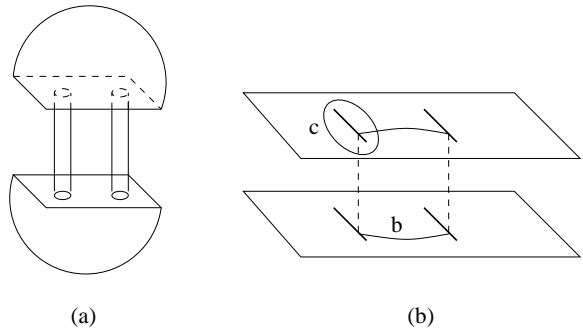}}
\caption{(a) Double sheeted cover of $x$-plane with branch cuts as a torus.
(b) $c$-type and $b$-type cycles on the cut $x$-plane.}
\label{fig:donut} \end{figure}

The Seiberg-Witten curves for these theories are 
hyperelliptic, taking the form,
\begin{equation}
y^2=F_{2N_c}(x), \label{eq:hyperelliptic}\end{equation}
where $F_{2N_c}(x)$ is a polynomial of order $2N_c$ in $x$, and is also a function
of the moduli $u_k$, or equivalently vevs of the symmetric polynomials, 
$s_k=(-1)^k\,\sum_{i_1<\cdots<i_k} \phi_{i_1}\cdots \phi_{i_k}$, $k=2,\dots,N$.
For SU$(N_c)$ with $N_c$ flavors, the curve takes the form \cite{HO}\footnote{
Our convention for the dynamical scale is different than that of \cite{HO}, 
namely, $\Lambda^{2N_c-N_f}=4\Lambda_{HO}^{2N_c-N_f}$.}
\begin{equation}
y^2=\left(\sum_{k=0}^{N_c}s_k\,x^{N_c-k}+\Lambda^{N_c}\right)^2-4\Lambda^{N_c}
\prod_{j=1}^{N_c}(x+m_j). \label{eq:SUNSW}\end{equation}
The complex variable $y$ is defined on a double-sheeted cover of the $x$-plane.
There are branch points at the roots of $F_{2N_c}(x)$ 
which are joined in pairs to
form cuts.  Identifying the point at infinity, the hyperelliptic curve has the
topology of a genus $N_c-1$ surface, as illustrated in Figure~\ref{fig:donut}a.
There are correspondingly two sets of one-cycles on the Seiberg-Witten curve,
as illustrated in Figure~\ref{fig:donut}b on the $x$-plane.  These cycles
are labeled $c_i$ and $b_i$, where $i=1,\dots,N-1$.  
For each pair of cycles
there is a holomorphic one-form $\omega_i$ which is defined globally on the
curve.  For a hyperelliptic curve (\ref{eq:hyperelliptic}), a basis of the
one-forms can be written as \cite{fulton}, \begin{equation}
\omega_i = \frac{dx\,x^{N_c-i}}{y(x)}, \ \ \ i=2,\dots,N_c. \end{equation}
As mentioned earlier, the gauge theory at strong coupling has a weakly 
coupled description in terms of a dual adjoint field $\Phi_D$. The dual
field $\Phi_D$ can be thought of as a particular
combination of the ``electric'' field
$\Phi$ and the ``magnetic'' field $\Phi_d$.  In terms of the vacuum expectation
values $a^i$ and $a_d^i$
of the diagonalized fields $\Phi$ and $\Phi_d$ the spectrum of
dyons with electric and magnetic charges $n_e^i$ and $n_m^i$ (and not
carrying global U(1) charges broken by the hypermultiplet masses $m_i$) 
is given by, \begin{equation}
M_{nm}= \left|\sum_i(n_e^i a^i+ n_m^i a_d^i)\right|. \label{eq:monmass}
\end{equation}
The vevs $a_i$ and $a_d^i$ are determined as functions of the moduli up to
gauge transformations via the identification \cite{AF,HO}, \begin{eqnarray}
\frac{\partial a^i}{\partial s_k}&=&\int_{c_i} \omega_k \\
\frac{\partial a_d^i}{\partial s_k}&=& \int_{b_i} \omega_k. \end{eqnarray}

Integrating these equations with respect to $s_k$ and using the weak
coupling vevs determines $a^i$ and $a^k$.  Equivalently, one can define
a meromorphic one-form (holomorphic up to a simple pole),
the Seiberg-Witten one-form $\lambda_{SW}$, such that the vevs are given by
\cite{AF,HO},\begin{eqnarray}
a^i&=&\int_{c_i} \lambda_{SW} \label{eq:a}\\
a_d^i&=& \int_{b_i} \lambda_{SW}. \label{eq:ad}\end{eqnarray}
The dyon masses (\ref{eq:monmass}) are then,\begin{equation}
M_{nm}=\left|\int_{n_e c_i+n_m b_i}\lambda_{SW}\right|.
\end{equation}
A dyon
becomes massless when a cycle of the Seiberg-Witten curve vanishes.
(A similar story applies when quarks become massless, but the masses
(\ref{eq:monmass}) contain a term proportional to the mass of the quark
and the global U(1) charge broken by the quark mass \cite{SW2}.)
This happens when two of the roots of $F_{2N_c}(x)$ coincide, as is evident
from Fig.~\ref{fig:donut}b.  When two roots collide either a branch cut
shrinks to zero size or two branch cuts collide.  Either way, a cycle
vanishes.  The degrees of freedom which become massless at
a given singularity are determined by the beta function of the weakly coupled
theory.  The beta function is calculated by studying how the period matrix,
\begin{equation}
\tau_{ij}\equiv\frac{\partial a_d^i}{\partial a^j}, \end{equation}
determined by (\ref{eq:a}) and (\ref{eq:ad}),
transforms when the moduli make a loop around the singularity.  This monodromy
depends on the beta function
because according to the Seiberg-Witten analysis $\tau_{ij}$ is identified
with the (running) gauge coupling function which appears in the action on
the Coulomb branch,
\begin{equation}
S\sim {\rm Im} \int d^4x \int d^2\theta \,\tau_{ij} \,W^{\alpha\,i}W_{\alpha}^j,
\end{equation}
with $W^{\alpha\,i}$ the field strength chiral superfield for the $i$th U(1)
gauge group factor.
If several quarks (under an unbroken SU$(r)$
factor in the gauge symmetry at a non-baryonic root) 
or mutually local monopoles 
(charged under orthogonal U(1) gauge group factors)
become massless, the roots of $F_{2N_c}(x)$ will coincide in
pairs, with one pair for each U(1) factor under which degrees of freedom 
become massless.  Higher order roots indicate
additional massless states, as in the presence of mutually nonlocal monopoles
at Argyres-Douglas points \cite{AD} or an enhanced gauge 
symmetry at a non-baryonic root in the theory 
with specially tuned quark masses.

At the ${\cal N}$=1 vacua discussed in the previous sections
$N_c-1$ mutually local monopoles become massless and condense at the baryonic 
branch, and on the non-baryonic branches $N_c-r$ monopoles (charged under 
$N_c-r$ unbroken U(1) factors) condense and
additional quarks become massless under the remaining unbroken
subgroup of SU$(r)$.  In either case, the
Seiberg-Witten curve is expected to factorize as,
\begin{equation}
y^2=f_2(x)\,\prod_{i=1}^{N_c-1}(x-x_i)^2, \label{eq:curvefact}\end{equation}
where $f_2(x)$ is a second order polynomial in $x$.
So the matrix model provides a prediction for the values of the moduli
$s_k$ for which the polynomial $F_{2N_c}(x)$ 
factorizes as (\ref{eq:curvefact}).
This is in general a complicated algebraic problem.  For the pure gauge theory
(without matter) there is a known solution in terms of Chebyshev polynomials
\cite{DS}.  In certain limits factorization of Seiberg-Witten curves has been
studied in theories with matter \cite{Murayama,Matt}.  Only recently has a
more general study been done via the matrix model duals of
the broken ${\cal N}$=2 gauge theories \cite{DJ,Demasure}, as reviewed
in Section~\ref{sec:Nf=Nc} for the $SU(N_c)$ gauge theory with 
$N_f<N_c$, and also as extended to the case 
$N_f$=$N_c$ there.  
The explicit predictions for curve factorization have been tested
in limited cases \cite{DJ,KS}.  In general the algebraic problem of 
factorization is quite complicated and has evaded an analytical solution.
In this section we study the results for
SU(2) with 2 flavors and SU(3) with 3 flavors in more detail, and we explicitly
test factorization of the Seiberg-Witten curves in that case.  In the case that
quark masses are chosen to be pairwise equal we will be able to classify 
the branches of moduli space in a simple way, as will be apparent from the 
M5-brane construction.
\subsection{SU(2) with 2 flavors}\label{sec:su2}
For simplicity we will take $m_1=m_2\equiv m$.  In this case there is a non-baryonic
Higgs branch in the ${\cal N}$=2 theory.  After breaking to ${\cal N}$=1 with the
adjoint mass the quantum modified 
constraint (\ref{eq:QMC})
on the diagonalized meson vevs is, \begin{equation}
\Xt_1\Xt_2=\Lambda^2. \end{equation}
Then the effective superpotential (\ref{eq:Weffmatrix}) can be written in terms
of $\Xt_1\equiv \Xt$ as, \begin{equation}
W_{eff}(\Xt)/M=
-\frac{1}{2}\left(\Xt^2+\frac{\Lambda^4}{\Xt^2}\right)+\frac{1}{4}
\left(\Xt+\frac{\Lambda^2}{\Xt}\right)^2+m\left(\Xt
+\frac{\Lambda^2}{\Xt}\right).\end{equation}
The stationary points of $W_{eff}$ are at the meson vevs: \begin{equation}\label{EQ357}
\Xt=\pm \Lambda,\ m\pm\sqrt{m^2-\Lambda^2}.
\end{equation}
These solutions give rise to the following 
values of the effective superpotential at
the vacua:
\begin{equation}
W_{eff}^{vac}/M=u_2^0=\pm 2m\Lambda\ {\rm or}\ \Lambda ^2+m^2. 
\label{eq:SU2Weff}\end{equation}

The last solution above is doubly degenerate (corresponding to
exchanging $\Xt_1$ and $\Xt_2$) and hence there are four 
vacua for this theory. 
These solutions are the matrix model and field theory prediction for 
$u_2$ at the vacua of the ${\cal N}$=2 theory broken to ${\cal N}$=1 by an
adjoint mass.

One can also obtain the effective superpotential directly from  matrix model. The 
baryonic $(r=0)$ solutions for R and T  are given by,
\begin{eqnarray}
2 T &=& (m + T - \sqrt{ (m+T)^2 -4R}) \\
\Lambda^2 &=& (\frac{1}{2}(m + T - \sqrt{ (m+T)^2 -4R}))^2,
\end{eqnarray}
which has the solutions $R= -\Lambda m, \,\, T= -\Lambda$ and  
$R= \Lambda m, \,\, T= \Lambda$. Substituting these values into (\ref{EQu2}) 
produces the first set of effective 
potentials in (\ref{eq:SU2Weff}). The equation of motion in the non-baryonic $(r=1)$
branch are,
\begin{eqnarray}
2 T &=& (m + T) \\
\Lambda^2 &=& R
\end{eqnarray} 
which yields the second set of solutions in   (\ref{eq:SU2Weff}).

For SU(2) the Seiberg-Witten curve (\ref{eq:SUNSW}) is,

\begin{equation}\label{eq:SU2SW}
y^2=\left(x^2-u_2+\Lambda^2\right)^2-4\Lambda^{2}(x+m_1)(x+m_2). 
\end{equation}
When the Seiberg-Witten curve factorizes it takes the form,
\begin{equation}
y^2=(x-e)^2(x^2+bx+c). 
\label{eq:SU2factcurve}
\end{equation}
We equate (\ref{eq:SU2SW}) and (\ref{eq:SU2factcurve}) to solve for the
factorized form of the curve and the value of $u_2$ at which the
curve factorizes.
For $m_1=m_2=m$ the solutions are, 
\begin{eqnarray}\label{EQ361}
u_2=m^2+\Lambda^2,&\  & y^2=(x+m)^2((x-m)^{2}-4\Lambda^2), \ {\rm or}
\nonumber\\
u_2=\pm 2m\Lambda&\  &y^2=(x\pm \Lambda)^2
\left(x^2-2\Lambda\,x+\Lambda^2 \mp 4m\Lambda\right).
\label{eq:SU2sols}\end{eqnarray}
These values for $u_2$ are in agreement with the matrix model prediction 
(\ref{eq:SU2Weff}).  The first solution in (\ref{eq:SU2sols}) corresponds to
the non-baryonic root, and the second to the 
baryonic root.

We also checked agreement between the various descriptions in the case 
$m_1=-m_2=m$, for
which there is a baryonic Higgs branch in the ${\cal N}$=2 theory.  Allowing for
a baryon vev, the effective superpotential after using the quantum modified constraint
is, \begin{equation}
W_{eff}(\Xt)/M=
-\frac{1}{2}\left(\Xt^2+\frac{(\Lambda^2+B\bar{B})^2}{\Xt^2}\right)+\frac{1}{4}
\left(\Xt+\frac{\Lambda^2+B\bar{B}}{\Xt}\right)^2+m\left(\Xt
-\frac{\Lambda^2+B\bar{B}}{\Xt}\right),
\end{equation}
where $B$ and $\bar{B}$ have been 
rescaled by the adjoint mass $M$ as $\Xt$ was.
Solving the equations of motion for $\Xt$ we find that at the non-baryonic 
vacua (for which 
$B=\bar{B}=0$), $W_{eff}=\Lambda^2\pm 2im\Lambda$, and
at the baryonic roots we find
$W_{eff}=m^2$, independent of $B$ and $\bar{B}$ as we have argued must be the case.

\subsection{SU(3) with 3 flavors}
\label{sec:su3}

The quantum modified 
constraint (\ref{eq:QMC})
on the diagonalized meson vevs is, 

\begin{equation}
\Xt_1\Xt_2\Xt_3=\Lambda^3. 
\end{equation}

Then the effective superpotential (\ref{eq:Weffmatrix}) can be written in terms
of mesons and a Lagrange  multiplier $\lambda_X$ as,

\begin{equation}
W_{eff}(\Xt)/M=
-\frac{1}{2}Tr (\Xt^2) + \frac{1}{6} (Tr \Xt)^2 + Tr m\Xt + \lambda_X ( \det
\,\Xt -\Lambda^3)  
\end{equation}
At the baryonic solutions the meson vevs are all equal. The vevs are
determined by the stationary points of $W_{eff}$ and are found to be,     
\begin{equation}\label{EQ364}
\Xt_i=\Lambda \omega  \qquad  i=1,2,3
\end{equation}
where  $\omega$ is a third root of unity.
At the remaining (non-baryonic) solutions two of three meson 
vevs are equal, namely,
\begin{eqnarray} 
\Xt_1 &=& \Xt_2= m - \frac{m^2}{\Delta}  -\Delta, \qquad \Xt_3= \frac{3m -
\Xt_1}{2} =
m + \frac{m^2}{2 \Delta}  + \frac{\Delta}{2}  \label{EQ365}\\
\Xt_1 &=& \Xt_2= m - \omega^2 \frac{m^2}{\Delta}  - \omega \Delta, 
\qquad \Xt_3 = m + \omega^2 \frac{m^2}{2\Delta}  + \frac{\omega \Delta}{2}   \label{EQ366}
\\
\Xt_1 &=& \Xt_2= m - \omega \frac{m^2}{\Delta}  - \omega^2 \Delta, 
\qquad \Xt_3= m + \omega \frac{m^2}{2\Delta}  + \frac{\omega^2 \Delta}{2}  \label{EQ367}
\end{eqnarray}  
where 
\begin{equation}
\Delta=  \left( \Lambda^3 - m^3 + \sqrt{ \Lambda^6 - 2 \Lambda^3 m^3} \right)
^{1/3}\label{eq:Delta}
\end{equation}
The first set of solutions  gives rise to the following 
values of the effective superpotential at
the three vacua:
\begin{equation}
W_{eff}^{vac}/M=u_2= 3m(\Lambda^3)^{1/3} . 
\label{eq:SU3Weff1}
\end{equation}
The second set of solutions gives rise to three vacua, one of which has
superpotential,

\begin{equation}
W_{eff}^{vac}/M=u_2= \frac{3}{4}( m^2 + 4 m \Xt_1 - \Xt_1^2)=   \frac{3}{4}
\left( 2 m^2 - \frac{m^4}{\Delta^2} -  \frac{2m^3}{\Delta} - 2 m \Delta - 
\Delta^2 \right).
\label{eq:SU3Weff2}
\end{equation}
The superpotential at the other two vacua can be obtained by replacing $\Delta$ with  
$\omega \Delta$ and $\omega^2 \Delta$. Notice that the first solution 
is invariant under  $\Delta \leftrightarrow \frac{m^2} {\Delta} $. In fact, 
each of these three solutions is three-fold degenerate (corresponding to
permutations of the three meson vevs). 
Thus, there are a total of twelve vacua for this theory.

Similarly,  the effective superpotential can be obtained from   matrix model. 
The equations determining $R$ and $T$ in the baryonic $(r=0)$ vacua are,
\begin{eqnarray}
3 T &=& \frac{3}{2} (m + T - \sqrt{ (m+T)^2 -4R}) \\
\Lambda^3 &=& \left(\frac{1}{2}(m + T - \sqrt{ (m+T)^2 -4R})\right)^3,
\end{eqnarray}
which have the solutions $R= \Lambda m, \,\, T= \Lambda$,   
$R= \Lambda \omega m , \,\, T= \Lambda \omega$ and $R= \Lambda \omega^2 m , 
\,\, T= \Lambda \omega^2$. Substituting these values into (\ref{EQu2}) 
produce the first set of effective 
potential in (\ref{eq:SU3Weff1}). One can also  obtain the moduli $u_3$ and 
in this branch, it is given by 
\begin{equation}
u_3= 3m (\Lambda \omega^k)^2  \,\;\;\; k=1 \ldots 3 
\label{eq:u3}
\end{equation}

The $R$ and $T$ equations for the $(r=1)$ vacua are,
\begin{eqnarray}
3 T &=& (m + T) + \frac{1}{2}(m + T - \sqrt{ (m+T)^2 -4R}) \\
\Lambda^3 &=& R \left(   \frac{1}{2}(m + T - \sqrt{ (m+T)^2 -4R})\right),
\end{eqnarray} 
which has the solutions,
\begin{eqnarray}
T &=& m-\frac{m^2}{2\Delta}- \frac{\Delta}{2} \\
R &=& -\frac{1}{2}\left(\, \frac{m^4}{\Delta^2} + \frac{m^3}{\Delta}
+ m \Delta + \Delta^2\right).
\end{eqnarray} 
Substituting to  (\ref{EQu2}), we find the second set of solutions 
(\ref{eq:SU3Weff2}). By replacing $\Delta$ with $\Delta \omega$ and  $\Delta \omega^2 $, we 
obtain the other two vacua. The modulus
$u_3$ is given in this vacuum by, \begin{equation}
u_3=\Lambda^3+m(u_2-m^2), \label{eq:u3NB}\end{equation}
with $u_2$ as given by (\ref{eq:SU3Weff2}).

For $SU(3)$ with three flavors of equal mass, the Seiberg-Witten 
curve (\ref{eq:SUNSW}) can be written as, 

\begin{equation}\label{eq:SU3SW}
y^2=\left(x^3-u_2 x -u_3 +\Lambda^3\right)^2-4\Lambda^{3}(x+m)^3 
\end{equation}
The factorized curve  takes the form,
\begin{equation}
y^2=(x-e_1)^2 (x-e_2)^2 (x^2+bx+c). 
\label{eq:SU3factcurve}
\end{equation}
By matching  (\ref{eq:SU3SW}) and (\ref{eq:SU3factcurve}), we  find the value 
 of $u_2$ and $u_3$ at which the curve factorizes.  With $u_2$ and
$u_3$  given by the baryonic
solution
(\ref{eq:SU3Weff1}), (\ref{eq:u3}) 
as predicted by the field theory and matrix model, the
curve indeed factorizes as,

\begin{multline}
y^2=\left(x + \frac{1}{2}(\Lambda + 
\sqrt{\Lambda(-3 \Lambda + 4m)} \right)^2  \left(x + \frac{1}{2}(\Lambda - 
\sqrt{\Lambda(-3 \Lambda + 4m)}\right)^2 \\ \times
\left(x^2-2\Lambda\,x+ \Lambda^2 - 4m \Lambda\right).
\label{eq:SU3solsB}
\end{multline}

At the non-baryonic solution given by (\ref{eq:SU3Weff2}) and (\ref{eq:u3NB}), 
the Seiberg-Witten curve again factorizes as (\ref{eq:SU3factcurve}) with:
\begin{eqnarray}
e_1&=&-m \nonumber \\
e_2&=& \frac{\Delta}{2} + \frac{m^2}{2\Delta} \nonumber \\
b &=& -2m+\Delta+\frac{m^2}{\Delta} \label{eq:SU3solsNB} \\
c&=& \frac{1}{e_2^2}\left((u_2-m^2)^2-4\Lambda^3m\right), \nonumber \end{eqnarray}
with $\Delta$ given by (\ref{eq:Delta}) and $u_2$ given by (\ref{eq:SU3Weff2}), in addition
to solutions of the same form with $\Delta\rightarrow \Delta \omega$ and $\Delta\rightarrow
\Delta\omega^2$.

As another consistency check on these solutions, 
consider the limit in which we keep $m\Lambda$ fixed while taking $\frac{m}
{\Lambda}$ large. The mesons are heavy with respect to the strong scale of 
the theory. In this case, the low energy theory reduces to 
${\cal N}=1$ $SU(N)$ SYM  gauge 
theory with no matter. This theory has $N$ distinct vacua located at 
$M u_2=N (\Lambda_{low}^{3N})^{1/N}$, with $\Lambda_{low}^3=Mm\Lambda$ from
the one-loop matching condition. 
A quick glance at the above examples for $SU(2)$ and 
$SU(3)$ gauge groups makes it clear that both the matrix model and 
curve factorization results are consistent with the pure 
gauge theory in the appropriate limit.  

Before we leave this section we make an observation regarding some of these solutions
that are indicative of a general result that we will prove in the following section.
In the SU(2) theory with equal quark masses, 
the first factorized form of the Seiberg-Witten curve in
Eq.~(\ref{EQ361}) has the form of $(x+m)^{2}$ times a polynomial whose interpretation is
{\em a priori} unclear.  In the SU(3) case with equal quark masses, 
the Seiberg-Witten curve factors similarly at
the non-baryonic vacua.  By insisting that the Seiberg-Witten curve factorize as,
\[
y^{2} = (x+m)^{2} f_4(x,u_{2},m),
\]
we find
\begin{equation}
    f_4(x,u_{2},m) = (x^{2}-m\, x - u_{2}+m^{2})^{2}-4 \Lambda^{3} (x+m)
\end{equation}
which describes the curve for the U(2) theory  with
$\nf=1$, with U(1) adjoint vev $\tilde{u}_1=m$ and $\tilde{u}_2=u_2-m^2/2$.  
This is not a coincidence, as we demonstrate in the following section.

\section{Relations between vacua}
\label{sec:BvsNB}
\setcounter{equation}{0}
In this section we test the decoupling of the Higgs and Coulomb branches, as
predicted by an ${\cal N}$=2 nonrenormalization theorem \cite{APS}, via
the matrix model solutions and Seiberg-Witten curves.  We also discuss the
persistence of ${\cal N}$=2 nonrenormalization properties in the
${\cal N}$=1 theories discussed previously.

\subsection{Relations between matrix model solutions}
On non-baryonic Higgs branches of the ${\cal N}$=2 SU$(N_c)$ theory
with $N_f$ flavors, which we will call the $(N_c,N_f)$ theory,
some of the adjoint vevs
$\phi_i$ become equal in magnitude to quark masses $m_i$.  As mentioned
earlier, the $D$-term and $F$-term equations of motion allow such solutions
to exist only when pairs of quark flavors have equal masses.  In the 
${\cal N}$=2 theory the squark vevs can be made arbitrarily large, and by the 
${\cal N}$=2 
nonrenormalization theorem the squark vevs cannot influence the Coulomb
branch of the remaining SU$(N_c-r)$ theory with $N_f-2r$ flavors (if
$r$ pairs of flavors obtain nonbaryonic vevs in this way) \cite{APS}.  
In particular, the location
on the Coulomb branch where monopoles become massless in the 
$r$th nonbaryonic branch of the ${\cal N}$=2
$(N_c,N_f)$ theory should be related to the effective superpotential at 
the vacua of the ${\cal N}$=1 $(N_c-r,N_f-2r)$ theory.
The matrix model solutions allow us to test this picture.

For simplicity we will first consider the case when all quark masses vanish, 
and afterward discuss the massive theory.  In the $r$th nonbaryonic
branch we can rewrite equations (\ref{EQu1}) and (\ref{EQLambda}) with
$n_-=r$ as,
\begin{eqnarray}
u_1(R,T)&=&\left(N_c-\frac{N_f}{2}\right)\,T-\left(\frac{N_f}{2}-r\right)
\sqrt{T^2-4R}, \\
\Lambda^{2N_c-N_f}&=&R^{N_c-N_f+r}\,\left(\frac{1}{2}(T+\sqrt{
T^2-4R})\right)^{N_f-2r}. \end{eqnarray}
Defining $\tilde{N}_c\equiv N_c-r$ and $\tilde{N}_f\equiv N_f-2r$, we then 
have, \begin{eqnarray}
u_1(R,T)&=&\left(\Nt_c-\frac{\Nt_f}{2}\right)\,T-\frac{\Nt_f}{2}
\sqrt{T^2-4R}, \\
\Lambda^{2\Nt_c-\Nt_f}&=&R^{\Nt_c-\Nt_f}\left(\frac{1}{2}(T+\sqrt{
T^2-4R})\right)^{\Nt_f}. \end{eqnarray}
These are the equations which define $R$ and $T$ in the baryonic solution of
the SU$(\Nt_c)$ theory with $\Nt_f$ flavors.\footnote{In fact, with $u_1\neq 0$
these equations also relate the vacua of the
U$(N_c)$ and U$(\Nt_c)$ theories, as is also expected from the ${\cal N}$=2
nonrenormalization theorem.}  
We can now demonstrate that
$u_2^0$ in the nonbaryonic $(N_c,N_f)$ vacua with massless flavors
equals $u_2^0$ in the vacua of the $(\Nt_c,\Nt_f)$ theory.  From (\ref{EQu2})
we have, \begin{eqnarray}
u_2^0&=&\left(\frac{T^2}{2}+R\right)\,\left(N_c-\frac{N_f}{2}\right)
+(N_f-2r)(-T)\sqrt{T^2-4R} \nonumber \\
&=&\left(\frac{T^2}{2}+R\right)\,\left(\Nt_c-\frac{\Nt_f}{2}\right)
+\Nt_f(-T)\sqrt{T^2-4R}. \end{eqnarray}
Note that in the above story we could have exchanged all positive and negative
square roots
and the conclusion would have been the same.
As we already mentioned, the $r$th set of 
nonbaryonic solutions corresponds to choosing either $r$ positive or $r$ negative
square roots, and the baryonic vacua correspond to choosing
either all positive or all negative square roots.  In either case, we have
shown that the values of $u_2^0$ agree in the corresponding baryonic and
nonbaryonic solutions.

Alternatively, we could have interpreted the above observation 
as a relation between vacua on the 
$(r-n)$th non-baryonic branch of the $(N_c,N_f)$ theory with the
$n$th non-baryonic branch of the $(N_c-(r-n),N_f-2(r-n))$th theory.
Hence, more generally, the value of the Coulomb branch modulus $u_2$ 
at the vacua in which monopoles are massless
on the $r$th Higgs branch of the ${\cal N}$=2
$(N_c,N_f)$ theory are equivalent to
the effective superpotential at the vacua of the ${\cal N}$=1 $(N_c-r,N_f-2r)$
theory, as promised.

This suggests that the Seiberg-Witten curve of the $(N_c,N_f)$ theory
at the $r$-th root factorizes in the following way (see also \cite{Murayama}):
\begin{equation}\label{EQconj1}
y^2 = x^{2 r} [ P_{\ncbar} ( x, \ut_k )^2 - 4 \Lambda^{2 \ncbar -\nfbar}
   x^{\nfbar}],
\end{equation}
where $P_{\ncbar}(x,\ut_k) = \sum_{k=0}^\ncbar \tilde{s}_k
x^{\ncbar-k}$. Note that
\[
y^2 = P_{\ncbar} ( x, \ut_k )^2 - 4 \Lambda^{2 \ncbar -\nfbar}
   x^{\nfbar}
\]
is the Seiberg-Witten curve of the $(\ncbar,\nfbar)$ theory. 
(To compare with earlier literature, 
this
factorization corresponds to the type I superconformal
theory of~\cite{Eguchi,Murayama}.\footnote{The 
$r=N_f/2$ case is an exception. It belongs to the
  class 3 or 4 SCFT, depending on whether $N_c-N_f/2$ is odd or
  even~\cite{Murayama}.})
Expanding (\ref{EQconj1}),
\begin{eqnarray*}
  x^{2 r} [ P_{\ncbar} ( x, \ut_k )^2 - 4 \Lambda^{2 \ncbar -\nfbar}
  x^{\nfbar}] & = & x^{2 r} [ (x^{\ncbar} - \ut_2 x^{\ncbar - 2} - u_3
  x^{\ncbar - 3} \cdots )^2 - \Lambda^{2 \ncbar-\nfbar} x^\nfbar ]\\
  & = & x^{2 r} x^{2 \ncbar} - 2 x^{2 r} \ut_2 x^{\ncbar - 2} + \cdots\\
  & = & x^{2 \nc} - 2 \ut_2 x^{2 \nc - 2} + \cdots .
\end{eqnarray*}
We find some evidence that the Seiberg-Witten curve of the $(N_c,N_f)$ theory
factorizes as (\ref{EQconj1}) in the fact that $u_2=\tilde{u}_2$, 
in accordance with the ${\cal N}$=2
nonrenormalization theorem.  In
fact, all of the $u_k=\tilde{u}_k$, $k=2,\dots,\ncbar$ in the curve factorized
as (\ref{EQconj1}), and the same should follow from the matrix model 
solutions although we have only checked this for $u_2$.

The story is only slightly more involved for the massive theory.  On
the $r$th nonbaryonic branch $r$ elements of the diagonalized adjoint $\phi_i$
satisfy $\phi_i+m_i=0$ for $r$ of the quark masses $m_i$. Recall also that
each of these $m_i$ is the mass of a pair of quarks.
In matching the moduli in the two
theories as above, there are two additional differences to keep in mind 
between the
massive and massless theories.  
First of all, in the $(N_c,N_f)$ theory the adjoint is
traceless.  That implies that in the corresponding $(\Nt_c,\Nt_f)$ theory
the trace of the adjoint vev, $\tilde{u}_1$, is determined by, \begin{equation}
u_1=\tilde{u}_1-\sum_{i=1}^r m_i=0. \label{EQu1ut1}\end{equation}
We can either think of this as turning on a vev for an extra U(1) in the
$(\Nt_c,\Nt_f)$ theory, or as shifting the average hypermultiplet mass by
$\sum_{i=1}^r m_i/\Nt_f$.

The other thing to keep in mind is that, in addition to the effects of
the extra U(1) 
adjoint vev due to (\ref{EQu1}), 
$u_2=1/2\,{\rm Tr}\,\phi^2$ receives a contribution from
the vevs $\phi_i=m_i$, so that, \begin{equation}
u_2=\tilde{u}_2+\frac{1}{2}\sum_{i=1}^r m_i^2. \label{EQu2ut2}\end{equation}

Keeping in mind these differences from the massless case, 
we can check that the  moduli agree in the massive
case just as for the massless case.  
Consider a vacuum on the $r$th non-baryonic branch of the ${\cal N}$=2
theory.  After breaking to the ${\cal N}$=1 theory this corresponds to any
of the vacua in the $(r+n)$th branch with $r+n\leq [N_f/2]$.
For the equations defining $T$ and $R$
we find, \begin{eqnarray}
T\left(\Nt_c-\frac{1}{2}\Nt_f\right)&=&-\sum_{i=1}^{\Nt_f-n}\frac{1}{2}\sqrt{
(m_i+T)^2-4R}+\sum_{i=\Nt_f-n+1}^{\Nt_f}\frac{1}{2}\sqrt{
(m_i+T)^2-4R} \nonumber \\
&&+\frac{1}{2}\sum_{i=1}^{\Nt_f}m_i+\sum_{i=1}^r m_i, 
\label{eq:Tmassive}\\
\Lambda^{2\Nt_c-\Nt_f}&=&R^{\Nt_c-Nt_f}\prod_{i=1}^{\Nt_f-n}\frac{1}{2}
(m_i+T-\sqrt{(m_i+T)^2-4R}) \nonumber \\
&&\times \prod_{i=\Nt_f-n+1}^{\Nt_f}\frac{1}{2}
(m_i+T+\sqrt{(m_i+T)^2-4R}), \label{eq:Lmassive}\end{eqnarray}
where we have used the fact that $r$ pairs of equal
$m_i$ appear in opposite branches of the square roots, corresponding
to the $r$th non-baryonic branch of the ${\cal N}$=2 theory..
Comparing with (\ref{EQu1}), we see that the last term in (\ref{eq:Tmassive})
plays the role of  $\tilde{u}_1$, as expected.  Then we can write
(\ref{EQu2}) as, \begin{eqnarray}
u_2^0&=&\left(\frac{T^2}{2}+R\right)\left(\Nt_c-\frac{\Nt_f}{2}\right)-
\sum_{i=1}^{\Nt_f-n}\frac{1}{4}(m_i-T)\sqrt{(m_i+T)^2-4R} \nonumber \\
&&+
\sum_{i=\Nt_f-n+1}^{\Nt_f}\frac{1}{4}
(m_i-T)\sqrt{(m_i+T)^2-4R}+\frac{1}{2}\sum_{i=1}^r m_i^2 
\nonumber \\
&=&\tilde{u}_2^0+\frac{1}{2}\sum_{i=1}^r m_i^2, \end{eqnarray}
with $\tilde{u}_2^0$ evaluated at the solutions of the
$(\Nt_c,\Nt_f)$ theory as promised.  
As in the massless case, all positive and negative branches of the square
roots could have been exchanged with the same conclusion.

Hence, in the generic case in which there exist pairs of quarks with equal
mass, we have once again proved that the non-baryonic roots of the
${\cal N}$=2 
$(N_c,N_f)$ theory at which monopoles condense determine the low energy 
superpotential at
all of the vacua of the ${\cal N}$=1 $(\Nt_c,\Nt_f)$  theory.

In accordance with this relation between the $(\nc,\nf)$ and $(\ncbar,\nfbar)$
theories, the curve of $(\nc,\nf)$ theory at the $r$-th
root should factorize as
\begin{equation}\label{EQconj1m}
  y^2 = ( x + m_1 )^{2} (x + m_2)^2 \cdots (x + m_r)^2 [ P_{\ncbar} (
  x, \ut_k )^2 - 4 \Lambda^{\ncbar}\prod_{k=r+1}^{\nf} (x+m_k) ].
\end{equation}
Again, $y^2 = P_\ncbar ( x, \ut_k )^2 - 4
\Lambda^{\ncbar}\prod_{k=r+1}^{\nf} (x+m_k)$ is the curve for
$(\ncbar,\nfbar)$ theory with $\ut_1=\sum_{i=1}^r m_i$. Now, as in the
$m_i=0$ case, the coefficient of $x^{\nc-2}$ should be equal to $-2
u_2$ for consistency with the picture developed above. 
Indeed, by expansion of Eq.~(\ref{EQconj1m})
and using Eqs.~(\ref{EQu1ut1}) and~(\ref{EQu2ut2}), the coefficient of
$x^{\nc-2}$ is,
\begin{equation}
2(\ut_1^2-\ut_2^2)-4\ut_1\sum_{i=1}^r m_i + \sum_{i=1}^r m_i^2 +
4\sum_{i<j}m_i m_j = -2 u_2,
\end{equation}
as desired. We also note that the coefficient of $x^{\nc-1}$ is
$-\ut_1+\sum_{i=1}^r m_i$, which vanishes by Eq.~(\ref{EQu1ut1}).

These results highlight the power of the matrix model
techniques and confirm the ${\cal N}$=2 nonrenormalization
theorem in this context.  We would like to stress the interesting fact that,
because the Coulomb branch
moduli at the ${\cal N}$=1 vacua are stuck to specific values
determined by the ${\cal N}$=2 theory, the low energy superpotential of the
${\cal N}$=1 theory inherits the nonrenormalization properties of
the ${\cal N}$=2 theory.  That is why the decoupling of the Higgs
and Coulomb branches of ${\cal N}$=2 theories
has proven valuable for studying theories with only ${\cal N}$=1
supersymmetry.

\subsection{Counting the solutions}

The matrix model results can be used to reproduce the number of vacua of
the $\mathcal{N}=1$ theory with equal mass $m_i=m$, computed
in~\cite{Murayama} (see also \cite{Ohta}). 
We present two different methods. First, we
simply count the number of solutions of $u_1(R,T)=0$ and
Eq.~(\ref{EQLambda}). The second method utilizes the relation between
baryonic and non-baryonic branches discussed in the previous
section, and provides another check of that relation.

First of all, $u_1 = 0$ implies
\begin{equation}\label{EQu1eq0}
\sqrt{(m+T)^2-4 R} = \frac{(2\nc-\nf) T - m \nf}{2r - \nf}.
\end{equation}
Assuming  
$r\neq\nf/2$, we use this to reduce
Eq.~(\ref{EQLambda}) to the following form:
\begin{multline}\label{EQredLambda}
\Lambda^{2\nc - \nf} = R^{\nc-\nf} \left((m+T)- \frac{(2\nc-\nf) T - m
    \nf}{2r - \nf} \right)^r \\
\times \left((m+T)+ \frac{(2\nc-\nf) T - m \nf}{2r -
    \nf} \right)^{\nf-r}.
\end{multline}
According to Eq.~(\ref{EQu1eq0}), $R\sim T^2 + \cdots$ and thus the order Eq.~(\ref{EQredLambda}) in $T$ is $2\nc -\nf$.
Therefore, there are $2\nc -\nf$ solutions.

To get the total number of solutions, we must take in account two
things. There is a degeneracy in choosing the sign in front of the
square root for each flavor. This accounts for a factor of $2^\nf$.
Another one is the symmetry of Eq.~(\ref{EQredLambda}) under $r \to
\nf-r$. The effect of the symmetry will be dependent on whether $\nf$
is even or odd. In the even case, the solutions will come in pairs. In
the odd case, a half number of solutions of $T$ will not be consistent
with the sign choice which was lost when (\ref{EQu1eq0}) was squared
to determine $R$. Therefore, in any case, it cuts the number of
solutions by a half. We conclude now that the total number of
solutions is $(2 \nc - \nf)\cdot 2^{\nf-1}$~\cite{Murayama,Ohta}.

In the case when $\nf$ is even and $r = \nf /2$, we obtain $T=0$ and
Eq.~(\ref{EQredLambda}) reduces to $R^{\nc-\nf/2} = \Lambda^{2 \nc
  -nf}$. Again, we get $(2\nc -\nf)/2$ solutions and the counting is the
same as above.

In the previous section, we showed that the $r$-th non-baryonic
branches of the $(\nc,\nf)$ theory correspond to the baryonic branch of
$(\ncbar,\nfbar)$ theory. In particular, the Seiberg-Witten curve factorizes as
Eq.~(\ref{EQconj1m}). Now, $P_\ncbar ( x, \ut_k )^2 - 4
\Lambda^{\ncbar}\prod_{k=r+1}^{\nf} $ must be factorized as the $r = 0$
case of $(\ncbar,\nfbar)$ theory.  This has $2 \ncbar - \nfbar = 2
\nc-\nf$ solutions.

Now, once again, we note the symmetry under $r \rightarrow \nf - r$.
This restricts $r\leq [\nf/2]$.  Also, there is a degeneracy for each
$r$, in choosing $r$ $(x+m)$ factors out of $N_f$ masses which are each equal
to $m$.  
In the case of odd
$\nf$, the counting is now complete as
\[
(2 \nc-\nf) \sum_{r=0}^{(\nf-1)/2} {\nf \choose r} = (2\nc-\nf) \cdot 
2^{\nf-1}.
\]
When $\nf$ is even, the case $r=\nf/2$ is subtle. It has only
$(\nc-\nf/2)$ solutions. However, total number of solutions is once
again
\[
(2\nc-\nf) \sum_{r=0}^{(\nf-1)/2} + (\nc-\nf/2){\nf \choose \nf/2} = 
(2\nc-\nf) \cdot 2^{\nf-1}.
\]
In fact, as readers may have noticed, our counting is in essence the
same as in~\cite{Murayama}.

Our counting is consistent with the examples in Section~\ref{sec:su2}
and~\ref{sec:su3}.  In Eq.~(\ref{EQ357}), we discovered $(2\cdot 2-2)\cdot
2=4$
solutions.  It would seem that Eq.~(\ref{EQ361}) has only three solutions, but
$u_{2}=m^{2}+\Lambda^{2}$ is a doubly degenerate solution as we noted.
For SU(3), we expect $(2\cdot 3-3)\cdot 2^{2} = 12$ solutions, as we found in
Section~\ref{sec:su3}.

\section{M-theory construction}
\label{sec:M}
\begin{figure}[t]
\epsfxsize=6in
\centerline{\epsfbox{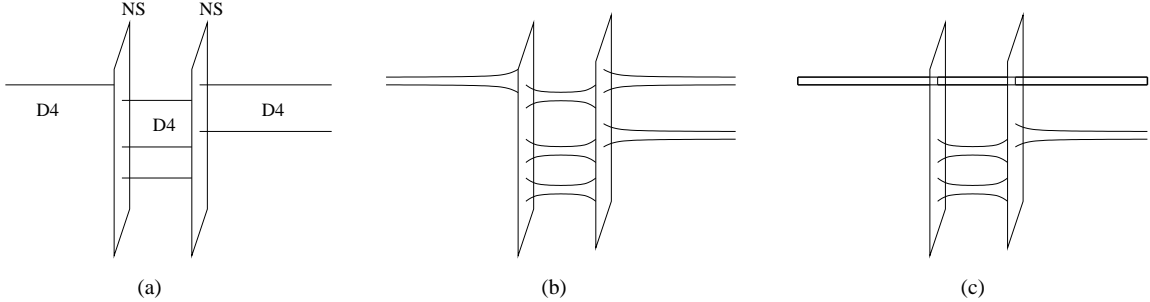}}
\caption{(a) Type IIA brane construction of ${\cal N}$=2 theory. (b) M5-brane
configuration of same theory. (c) Seiberg-Witten curve factorizes at non-baryonic
root of the Higgs branch.}
\label{fig:HW} \end{figure}

\setcounter{equation}{0}
As discussed in the Introduction, a non-baryonic Higgs branch opens up
when a pair of flavor masses become equal.  Furthermore, as discussed in 
Section~\ref{sec:BvsNB}, the effective superpotential in the 
non-baryonic vacua of the ${\cal N}$=1 theory
matches the baryonic vacua 
of the theory with one fewer colors and two fewer flavors.
We can easily understand this from the
M5-brane construction of these theories.  Witten has demonstrated that
the Seiberg-Witten curve of an ${\cal N}$=2 gauge 
theory can often be determined
as the geometry of the M5-brane configuration whose low energy dynamics is
described by the gauge theory \cite{Witten}.  The Type IIA brane configuration
for the ${\cal N}$=2 SU$(N)$ gauge theory with $N$ flavors
corresponds to a stack of $N$ D4-branes stretched between a pair of NS5-branes,
with $N$ semi-infinite D4-branes attached to each of the NS5-branes
\cite{HW}, as in Figure~\ref{fig:HW}a.  
The positions of the finite D4-branes correspond to the
adjoint vevs, and the positions of the semi-infinite D4-branes correspond to
the hypermultiplet masses.  
In the M-theory picture, the entire setup is 
described by a single M5-brane, in which the D4-branes correspond
to the M5-brane wrapped around the M-theory circle, and the NS5-branes
correspond to the M5-brane at a point on the M-theory circle, as in 
Fig.~\ref{fig:HW}b.  
We can set pairs of hypermultiplet masses
equal by matching the positions of the intersection of the
semi-infinite D4-branes with the NS5-branes on opposite NS5-branes.  In this
way, when one of the D4-branes suspended between the NS5-branes aligns itself
with that pair of semi-infinite D4-branes, they merge and become one large
D4-brane (Fig.~\ref{fig:HW}c)
which can then be removed from the rest of the brane configuration in a 
transverse direction
without any cost of energy.  This represents a non-baryonic root of the moduli
space.  
When the Coulomb and Higgs branches meet, the M5-brane
``tube'' can be removed, and the algebraic curve describing the M5-brane
factorizes.  An ${\cal N}$=2 nonrenormalization theorem implies that the Higgs
branch and the Coulomb branch are decoupled \cite{APS}, so
after factoring out the tube, what remains of the M5-brane geometry
must be the Seiberg-Witten
curve of the Higgsed theory with one fewer colors and two fewer flavors,
as we found evidence for in Section~\ref{sec:BvsNB}.  
The baryonic branch corresponds to the joining of each of the finite D4-branes with a
semi-infinite D4-brane, which because of the freezing of the U(1)
can only happen if the average position of both types of D4-branes is the same,
in which case the NS5-brane at the junction can be removed from the configuration.

\section{Conclusions}
\label{sec:Conclusions}
We have extended previous predictions for the
vacuum structure of broken ${\cal N}$=2 gauge theories with matter to the
case of SU$(N_c)$ gauge group with equal number of flavors and colors.  Below
the scale of the massive adjoint hypermultiplet this theory acquires a quantum
modified constraint on the moduli due to nonperturbative effects.  With this
constraint we determined the field theory expectation for the low energy
superpotential below the lowest mass scale of the theory.  We found a number
of solutions corresponding to the various vacua of the theory.  These results
were compared with the matrix model prediction for the low energy 
superpotential and were found to agree.  Although the presence of baryons in
the theory could potentially have been a problem for the naive generalization
of the matrix model results for $N_f<N_c$ to this case, we argued that the low energy
superpotential is independent of the baryon vevs.
We studied the
SU(2) and SU(3) theories in detail and verified the matrix model
prediction for the location on the Coulomb branch at which 
the Seiberg-Witten curve maximally factorizes,
with $(N_c-1)$ pairs of double roots.  
It would be interesting to extend
these results to theories in which the baryons play a nontrivial role, 
for example by adding a tree level superpotential depending on them.

We also demonstrated a relation
between the nonbaryonic vacua of the SU$(N_c)$ theory  with $N_f$ flavors
and specially tuned
masses and the vacua of the SU$(N_c-r)$ theory with $N_f-2r$ flavors.
To be precise, we checked that one of the moduli $(u_2)$ matches between these
vacua, and by the ${\cal N}$=2 nonrenormalization theorem we expect the same
to be true for the other moduli.  It would be nice to check this claim 
explicitly via the matrix model solutions.

\section*{Acknowledgments} 
We are happy to thank Yves Demasure, 
Tim Hollowood, Andreas Karch,  Yuri Shirman, Matt Strassler and 
John Terning 
for useful conversations.  JE also thanks the Aspen Center for Physics where
parts of this work were done. 
JE is supported by the DOE under contract 
DE-FGO2-96ER40956.
SH is supported by the DOE under contract DOE-FG02-95ER40893.
MU is supported in part by  DOE grant  DE-FGO3-00ER41132.

\vspace{\baselineskip}

\end{document}